\documentclass[11 pt,leqno]{article}
\usepackage{latexsym}
\usepackage{amssymb,amsbsy,amsmath,amsfonts,amscd}
\usepackage{vmargin}
\usepackage{graphicx}

 \usepackage{color}
\usepackage{mathenv}

\newenvironment{changemargin}[2]{\begin{list}{}{%
\setlength{\topsep}{0pt}%
\setlength{\leftmargin}{0pt}%
\setlength{\rightmargin}{0pt}%
\setlength{\listparindent}{\parindent}%
\setlength{\itemindent}{\parindent}%
\setlength{\parsep}{0pt plus 1pt}%
\addtolength{\leftmargin}{#1}%
\addtolength{\rightmargin}{#2}%
}\item }{\end{list}}

\newtheorem{lemme}{Lemme}[section]

\newtheorem{Theorem}{Theorem}[section]

\newtheorem{theorem}{Th\'eor\`eme}[section]

\newcommand{\Ind}{\ifmmode{{\rm 1} \hskip -3pt {\rm I}}
    \else{\hbox{$1\hskip -3pt {\rm I}$}}\fi}

\catcode`\@=11
\newbox\bz@
\newdimen\bdimz@
\def\linethrough#1{\setbox\bz@=\hbox{#1}%
\bdimz@=\ht\bz@ \divide\bdimz@ by 2 \advance\bdimz@ by -\dp\bz@ \ht\bz@=\bdimz@
\leavevmode\hbox{$\overline{\box\bz@}$\relax}}
\catcode`\@=12

\newcommand{\g}{\nabla}

\newcommand{\Om}{\Omega}
\newcommand{\G}{\Gamma}
\newcommand{ \vit}{\hbox{\bf u}}

\newcommand{ \vittest }{\hbox{\bf v}}

\newlength\jataille
\newcommand{\figgauche}[5]%
{\jataille=\textwidth\advance\jataille by -#1
\advance\jataille by -.5cm
\begin{figure}[!h]
\begin{minipage}[c]{#1}
\hskip-0.5cm \includegraphics[width=#1]{#2}\\
Figure 3.3: \small{\textit{Profil vertical de la vitesse horizontale. Mod\`ele de d\'econvolution en bleu (LES-1), mod\`ele Leray-$\alpha$ en noir (Leray-$\alpha$) et DNS en rouge}}
\label {#4}
\end{minipage}\hfill
\begin{minipage}[c]{\jataille}
\normalsize #5 \normalsize
\end{minipage}
\end{figure}}

\makeatletter
\renewcommand{\thefigure}{\ifnum \c@section>\z@ \thesection.\fi
 \@arabic\c@figure}
\@addtoreset{figure}{section}
\makeatother

\makeatletter
\@addtoreset{equation}{section}

 \usepackage{epsfig}

\newcommand{\E}{\varepsilon}
\newcommand{\p}{\partial}

\newcommand{\R}{\ifmmode{{\rm I} \hskip -2pt {\rm R}}
    \else{\hbox{$I\hskip -2pt R$}}\fi}

\newcommand{\N}{\ifmmode{{\rm I} \hskip -2pt {\rm N}}
    \else{\hbox{$I\hskip -2pt N$}}\fi}

\newcommand{\x}{{\bf x}}

\newcommand{\BEQ} {\begin{equation}}
\newcommand{\EEQ} {\end{equation}}
\newcommand{\BTHM} {\begin{Theorem}}
\newcommand{\ETHM  } {\end{Theorem}  }

\date{ }

\begin{document}

{\sl  \small Projet de note CRAS, calcul scientifique, soumise \`a O. Pironneau}

\bigskip
\begin{changemargin}{1cm}{1cm} { \begin{center}  \sc \large Simulations de l'\'ecoulement turbulent marin avec un mod\`ele de d\'econvolution\end{center}}  \end{changemargin}

\begin{center}Anne-Claire Bennis \footnote{IRMAR, Campus Beaulieu, Universit\'e de Rennes
I, 35000 RENNES, France,
Anne-Claire.Bennis@univ-rennes1.fr},  Roger Lewandowski \footnote{IRMAR,
Roger.Lewandowski@univ-rennes1.fr, http://perso.univ-rennes1.fr/roger.lewandowski/}   et Edriss S. Titi \footnote{Department of Computer Science and Applied Mathematics,
Weizmann Institute of Science
Rehovot, 76100, ISRAEL,
also Department of Mathematics and Department of Mechanical and Aerospace Engineering, University of California, Irvine, CA 92697-3879, USA} \end{center}

\begin{center} {\bf R\'esum\'e} \end{center}

\begin{changemargin}{1cm}{1cm} {\footnotesize  \hskip 0.7cm
On introduit une \'equation de d\'econvolution  qui g\'en\'eralise l'algorithme de Van-Cittert pour des conditions aux limites de type oc\'eananique avec vent fix\'e. On en d\'eduit un mod\`ele de SGE pour lequel on a existence et unicit\'e d'une solution r\'eguli\`ere. Nous d\'etaillons un ensemble de simulations num\'eriques qui montrent l'int\'er\^et pratique du mod\`ele.  } \end{changemargin}

 \begin{center} {\bf Abstract} \end{center}

\begin{changemargin}{1cm}{1cm} {\footnotesize  \hskip 0.7cm
We display a continous equation for  the deconvolution process that generalizes the Van Cittert algorithm in the case of oceanic boundary conditions for a given fixed wind.  We deduce a LES model for which we have existence and uniqueness of a strong solution. Finally, we display several numerical simulations showing the practical interest of the model. } \end{changemargin}
\medskip

{\bf \large Abridged English version}
\medskip

A LES model  based for the Navier-Stokes equations (NSE), where the convection term is replaced by $(H_N (\vit) \cdot \g) \vit$, was introduced recently in \cite{La08}. In this model, $H_N (\vit)$  depends on $\vit$ in the following way. For a given $\vit$, with $\g \cdot  \vit = 0$, we filter $\vit$ by solving the Stokes (Helmholtz) like elliptic problem $-\alpha^2 \Delta \overline \vit +  \overline \vit+  \g \pi = \vit$, $\g \cdot  \overline \vit = 0$ subject to the appropriate boundary conditions, and we then set $A \overline \vit = \vit$. We then follow the  Van Cittert's algorithm (see \cite{La08}), $\vit_0 = \overline \vit $,  $\vit_{n+1}= \vit_{n}+\{\overline{\vit}-A^{-1}\vit_{n}\}$ and we set
$ \vit_{N}=H_{N} (\vit) $ for some $N$. This model makes sense in the case of periodic boundary conditions (see \cite{La08}). Nevertheless, we cannot use it
 in the case of realistic nonhomogeneous boundary conditions, such as those satisfied by oceanic flows with a fixed wind stress at the surface under the rigid lid hypothesis. For this reason a substantial modification is necessary.

 Our contribution starts from  observing  that the Van Cittert algorithm is in fact nothing but a finite difference scheme of a  certain evolution equation.  Taking $\tau$ as the phony time evolution parameter, this equation for scalar quantities takes the form  $\displaystyle -\alpha^2 \Delta {\p v \over \p \tau } + v = u$, $v \vert_{\tau = 0} = \overline u$. In this work, we make use of this remark to study the case of an oceanic  flow satisfying the boundary conditions  $(\ref{BC})$, below. Here $\x = (x,y,z) \in \R^3$, ${\cal O}  \subset \R^2$ is a bounded smooth domain, $\Om = {\cal O}  \times [-h, 0]$, $ \Gamma_s = {\cal O}  \times \{0 \}$,
 $ \Gamma_l = \p {\cal O}  \times [-h, 0]$, $\Gamma_b = {\cal O}  \times \{-h  \}$;
$ \Gamma = \Gamma_s \cup \Gamma_l \cup \Gamma_b$. Finally $t \in [0,T]$ is the real time parameter. We first introduce the system $(\ref{Filter})$-$(\ref{EQSS})$. This yields the deconvolution model $(\ref{pb1})$.
In this model, we do not make the hydrostatic balance assumption in order to capture and model the vertical convection. For the sake of simplicity, we do not consider the Coriolis force, which can be added to $(\ref{pb1})$, without causing any difficulties in our analysis.
We now give a short description of our results that will be reported in detail in \cite{Le08}. The space functions are defined in $(\ref{ESP})$ below.  For a given wind stress ${\bf V}$, $\boldsymbol{\psi} = \boldsymbol{\psi}({\bf V})$ is a divergence free vector field  which satisfies the boundary conditions $(\ref{BC})$
and is defined by $(\ref{FFRT})$.   Let $\tau >0$ be given, and $\vit = \vit(t, \x)$ with $\g \cdot \vit =  0$. We set $H_\tau (\vit)(t, \x) =
\vittest (\tau, t, \x) + \boldsymbol{\psi} ({\bf V})(\x)$ where $\vittest (\tau, t, \x)$ is solution of  system $(\ref{Filter})$-$(\ref{EQSS})$ for a fixed time $t$ (see $(\ref{FORM})$ below). Notice that when $\tau=0$,
$H_0 (\vit) = \overline \vit$ and therefore our model is precisely the Leray-$\alpha$ model, as considered in \cite{CHOT05} and \cite{GH03}. When $\tau = N$ is an integer, this model remains the same as the one introduced in \cite{La08} for the periodic boundary conditions case,  and here it is adjusted to satisfy physical boundary conditions. Our main results are
\begin{Theorem}Ê\label{Beau} Assume ${\bf V} \in H^3({\cal O})^2$ with a compact support, $\vit_0  \in {\bf I \!  H}_1 $ (see definition $(\ref{ESP})$) and
${\bf f} \in
C^0([0,T],  {\bf I \!  H}_1)$.
Let $\tau >0$ be given, Problem $(\ref{pb1})$ has a unique strong solution
$(\vit_\tau, p_\tau) \in L^2([0,T], H^2(\Om)^3)
\times L^\infty([0,T], H^1(\Om))$ with $\vit_\tau \in C([0,T], L^2(\Om)^3)$,
$\p_t \vit_\tau \in L^2 ([0,T] \times \Om)^3$. \end{Theorem}

\begin{figure}[h!] \label{FIG1}
\hskip-2.4cm\includegraphics[scale=0.45]{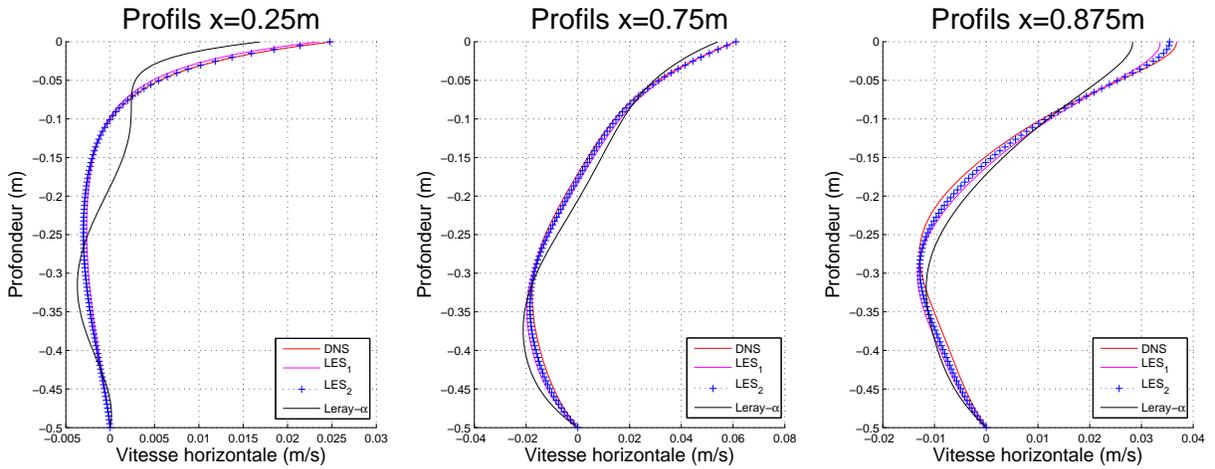}
\caption{\small{\textit{Profils verticaux de la vitesse horizontale pour la DNS (DNS), pour le mod\`ele de d\'econvolution avec $\tau=5$ (LES-1) et avec $\tau=20$ (LES-2) et pour le mod\`ele Leray-$\alpha$ (Leray-$\alpha$)}}}
\end{figure}

\begin{figure}[h!] \label{FIG2}
\begin{tabular}{ccc}
\hskip -1.1cm\includegraphics[scale=0.33]{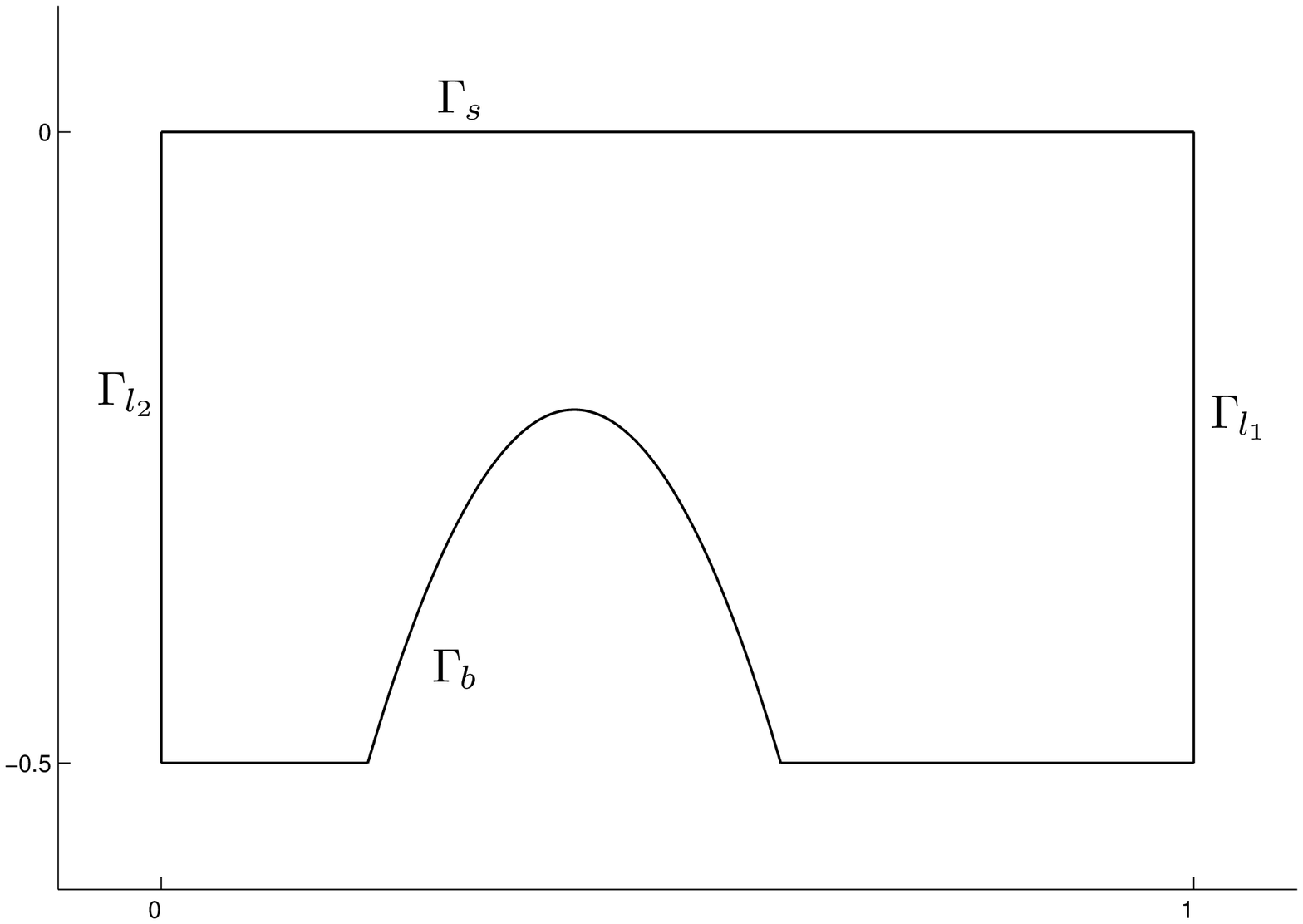}&&
\hskip0.35cm\includegraphics[scale=0.55]{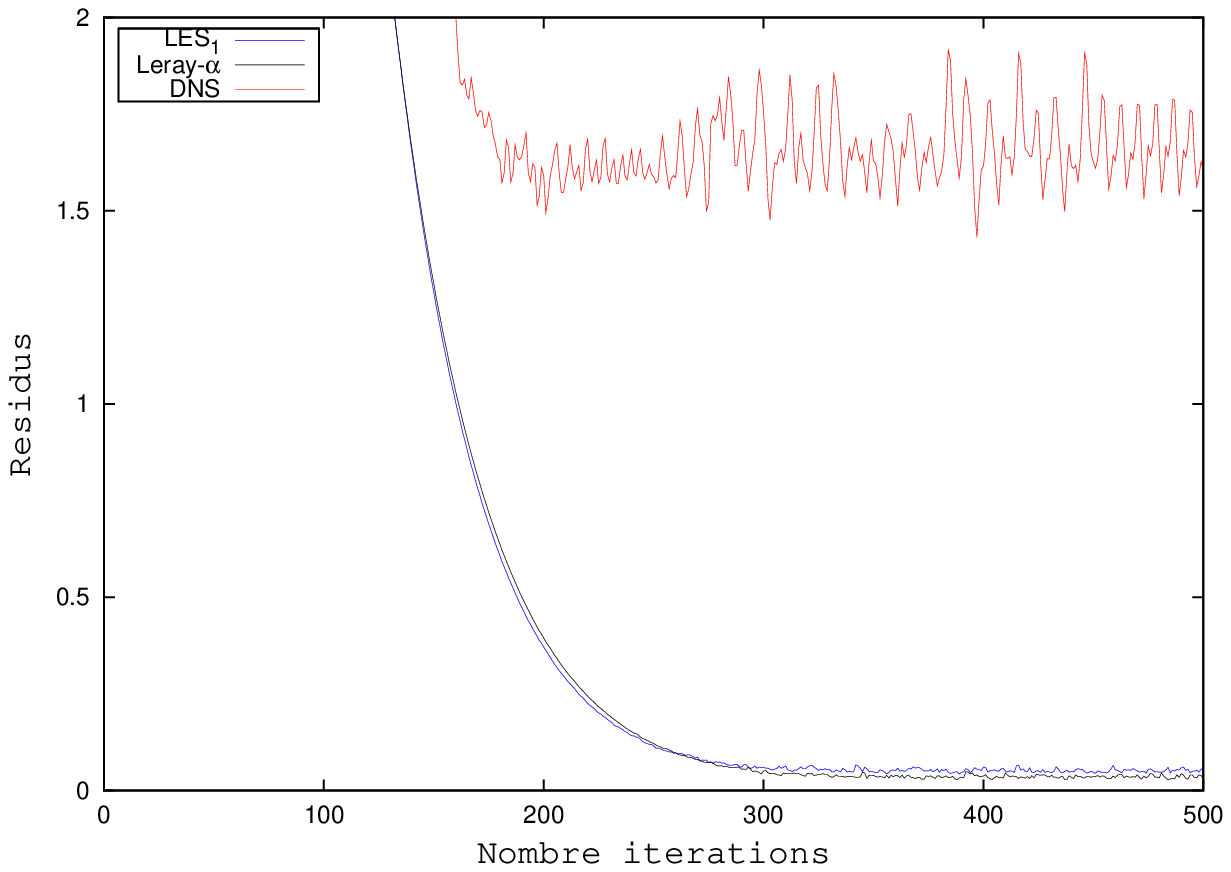}
\end{tabular}
\vskip-0.3cm
\caption{\small{\textit{G\'eom\'etrie du domaine \`a gauche et R\'esidus des \'equations de Navier-Stokes \`a droite (DNS en rouge, mod\`ele de d\'econvolution en bleu et mod\`ele Leray-$\alpha$ en noir)}}}
\end{figure}

\begin{Theorem} \label{FINAL}  There exists $(\tau_n)_{n \in N}$, which goes to infinity, when $n$ goes to infinity, and such that the sequence $(\vit_{\tau_n})_{n \in  \N}$ converges weakly in
$ L^2([0,T], {\bf I \!  H}_1)$ and strongly in  $ L^2([0,T], {\bf I \!  H}_0)$, to a field
$\vit = (\vit_h, w) \in L^2([0,T], {\bf I \!  H}_1) \cap L^\infty([0,T], L^2(\Om)^3)$, a Leray-Hopf weak solution to the NSE, with
 $\p_t \vit \in L^{4 \over 3}Ê([0,T],  {\bf I \!  H}_1')$ and such that for all
$\vittest = (\vittest_h, \theta) \in L^4([0,T], {\bf I \!  H}_1)$
$(\ref{varfaib})$ is satisfied. Moreover, $\vit (t, \cdot)$ converges weakly to $\vit_0$ in $L^2(\Om)^3$ as $t$ goes to $0$. Moreover, $\vit$ satisfies the energy inequality.
\end{Theorem}
Finally, we display numerical tests by using FreeFem++ \cite{He06} in the 2D case. Our code is semi-implicit and makes use of a method of the caracteristics adjusted to the deconvolution. Two cases have been considered. For each case, the wind stress is equal to ${\bf V}(x)=
a(x)\sin(\pi x)$, $\x = (x,y)$, $\x_h = x \in \G_s \approx [0, 1]$, where  $a(x)$ has a compact support and it localizes the stress. In the first case, $\Om$ is a square,
$\Delta t = 0.2 \, s$, $\alpha = 0.1 \, m$, $T = 90 \, s$. The DNS exhibits a good behavior. We compare the cases where the order of deconvolution $\tau$ takes for values $\tau=0$ (Leray-$\alpha$ model), $\tau = 5$ and $\tau=20$, $\Delta \tau = 1$. Vertical profils are displayed in Figure
\ref{FIG1}. They confirm the convergence Theorem and the validity of the code.
The second case is the one where $\Om$ has a topography (see Figure \ref{FIG2}),
$\Delta t = 0.02 \, s$, $\alpha = 0.1 \, m$, $T = 10 \, s$, $\tau = 5$, $\Delta \tau = 1$.  Poiseuille Flows have been imposed on the lateral boundaries to enhance turbulence and the DNS converges very slowly. Figure 3.3 displays one profil and
underlines the improvement of the computation using the model compared with the Leray-$\alpha$ model. Therefore, our model appears to produce good results and hence  could be suggested as a good alternative for simulations in real situations.

\section{Introduction}
\vskip -0,2 cm

On a introduit dans \cite{La08}  un mod\`ele de SGE bas\'e sur les \'equations de Navier-Stokes, dans lequel le terme de convection vaut $(H_N (\vit) \cdot \g) \vit$, $H_N (\vit)$ \'etant d\'eduit de la vitesse $\vit$ (avec $\g \cdot \vit = 0$) en partant de la filtration $-\alpha^2 \Delta \overline \vit +  \overline \vit+  \g \pi = \vit$, $\g \cdot \overline \vit = 0$. On pose $A \overline \vit = \vit$, que l'on d\'econvole \`a l'aide de l'algorithme de Van Cittert, $\vit_0 = \overline \vit $,  $\vit_{n+1}= \vit_{n}+\{\overline{\vit}-A^{-1}\vit_{n}\},
 \vit_{N}=H_{N} (\vit) $. Ceci a un sens dans le cas de conditions aux limites p\'eriodiques et permet d'approcher les \'equations de Navier-Stokes par des probl\`emes bien pos\'es avec un bon contr\^ole de l'erreur quand les lois de Kolmogorov sont satisfaites.

Il semble difficile d'appliquer cette proc\'edure telle quelle dans le cas de conditions aux limites r\'ealistes, comme par exemple pour l'Oc\'ean avec une tension du vent en surface fix\'ee, sous l'hypoth\`ese du to\^it rigide. La remarque principale dans ce travail est que l'algorithme de Van Cittert peut \^etre vu comme l'\'equation aux diff\'erences finies d'une \'equation d'\'evolution  qui s'\'ecrit pour des scalaires sous la forme $\displaystyle -\alpha^2 \Delta {\p v \over \p \tau } + v = u$, $u \vert_{\tau = 0} = \overline u$, o\`u $\tau$ est le param\`etre de d\'econvolution. Nous  exploitons cette remarque  de mani\`ere rigoureuse pour les conditions aux limites  $(\ref{BC})$ satisfaites par la vitesse de l'\'ecoulement marin,
 en introduisant le syst\`eme $(\ref{Filter})$-$(\ref{EQSS})$. Cela nous permet d'introduire le mod\`ele de SGE $(\ref{pb1})$. Dans ce mod\`ele, nous ne faisons  pas l'hypoth\`ese hydrostatique pour mieux d\'ecrire les effets convectifs et capter la couche de m\'elange. Pour simplifier la pr\'esentation, nous n'avons pas pris en compte la force de Coriolis, qui peut \^etre rajout\'ee dans $(\ref{pb1})$ sans changer la nature de notre analyse.

 Nous montrons que ce mod\`ele conduit \`a un probl\`eme bien pos\'e et nous avons un r\'esultat de convergence vers les \'equations de Navier-Stokes quand le param\`etre de d\'econvolution tends vers l'infini. Dans une s\'erie de tests num\'eriques, nous validons le mod\`ele dans le cas 2D en comparant les r\'esultats obtenus avec ceux d'une DNS convergente. Nous produisons des calculs num\'eriques dans un cas  avec topographie, cas pour lequel la DNS a un mauvais comportement. Le mod\`ele donne des bons r\'esultats num\'eriques pour $\tau =5$ qui affinent ceux donn\'es par les Leray-$\alpha$ mod\`eles habituels adapt\'es \`a nos conditions aux limites avec notre m\'ethode, puisqu'ils correspondent au cas $\tau=0$ (voir les mod\`eles d'origine dans \cite{CHOT05}, \cite{GH03}).
  Cette note annonce des r\'esultats d\'evelopp\'es dans l'article en cours de finalisation \cite{Le08}.

\section{Resultats th\'eoriques}

\vskip -0,2 cm
\subsection{G\'eom\'etrie et espaces fonctionnels}Ê
\vskip -0,5 cm

On note $\x = (x,y,z)$ un point de $\R^3$, ${\cal O}  \subset \R^2$ un domaine born\'e r\'egulier de $\R^2$, $\Om = {\cal O}  \times [-h, 0],$
o\`u $h>0$. Soient $\Gamma_s$ la surface,  $ \Gamma_s = {\cal O}  \times \{0 \}, $
et $\Gamma_l$ et $\Gamma_b$ les fronti\`eres lat\'erales et le fond,
$ \Gamma_l = \p {\cal O}  \times [-h, 0],  \quad  \Gamma_b = {\cal O}  \times \{-h  \}$.
On note enfin $ \Gamma = \Gamma_s \cup \Gamma_l \cup \Gamma_b$.

Soit $\vit = (\vit_h, w)$ la vitesse de l'eau de mer, $\vit_h = (u,v)$ sa partie horizontale. Les conditions aux limites classiques sont (sous l'hypoth\`ese du to\^it rigide)
\BEQ \label {BC}Êw\vert _\Gamma = 0, \quad {\vit_h} \vert_ {\Gamma_l \cup \Gamma_b }Ê= 0,
\quad \g {\vit_h} \cdot{\bf n} \vert _{\Gamma_s} = {\bf V},
\EEQ
o\`u ${\bf n}$ est la normale sortante, ${\bf V} = {\bf V}(x,y)$ est le vent (ici fix\'e) d\'efini sur ${\cal O}$. On identifie ${\cal O}$ et  $\G_s$. Les espaces fonctionnels sont les suivants :
\BEQ \label {ESP} \begin{array} {l} {\bf I \!  H}_0 = \{ \vit = (\vit_h, w) \in L^2(\Om)^3 ; \, \, \vit \,   \cdot{\bf n}Ê\vert_ {\Gamma}  = 0,  \,   \,
\g \cdot \vit = 0 \}, \cr
{\bf I \!  H}_1 = \{ \vit = (\vit_h, w) \in H^1 (\Om)^3, \, w \vert_{\Gamma}Ê= 0, \,  \vit_h \vert_ {\Gamma_l \cup \Gamma_b }Ê= {\bf 0},
 \g \cdot \vit = 0 \}, \end{array} \EEQ
 Nous montrons dans \cite{Le08} l'existence d'une base sp\'eciale adapt\'ee,  r\'esum\'e dans ce qui suit.
 \begin{lemme} \label{BSP} Il existe une suite de r\'eels strictement positifs $(\lambda_n)_{n \in \N}$ des suites $({\bf e}_n) _{n \in \N}$,  $({\pi}_n) _{n \in \N}$ o\`u ${\bf e}_n \in  {\bf I \!  H}_1 \cap H^2(\Om)^3 $, $\pi_n \in H^1(\Om)$ pour chaque $n$ et tels que
 $- \Delta {\bf e}_n + \g \pi_n = \lambda_n {\bf e}_n$.
 \end{lemme}
 La d\'emonstration utilise un argument de sym\'etrie pour se ramener un probl\`eme de Stokes sur un cylindre avec des conditions aux limites de Dirichlet homog\`enes. Un r\'esultat analogue avec des \'equations primitives a d\'ej\`a \'et\'e prouv\'e dans \cite{MZ952} avec la m\^eme technique. Nous montrons dans \cite{Le08}  que l'injection
 $  {\bf I \!  H}_1 \subset  {\bf I \!  H}_0$ est dense.
 Nous terminons cette section par le rel\`evement de la condition aux limites. On suppose que  ${\bf V} = {\bf V} (x,y)$ est \`a support compact dans  $\G_s$. Soit   $\boldsymbol{\psi} = \boldsymbol{\psi}(x,y,z)$ le champ de vecteurs d\'efini dans $\Om$ par
\BEQ \label{FFRT} \boldsymbol{\psi}(x,y,z)= \left(
\begin{array}{l}
\boldsymbol{\psi}_{h}  \\
\psi_{v}
\end{array} \right )  =
\left ( \begin{array}{l} \rho(z)\textbf{V}(x,y) \\ -\kappa(z)(\nabla_{h}\cdot\textbf{V})(x,y)
\end{array}
\right) =  \boldsymbol{\psi} ({\bf V}), \EEQ
o\`u
$\kappa(z)=\int^{z}_{-h}\rho(z^{'})dz^{'}$ et $\rho(z)=\frac{3}{4h}z^{2}+z+\frac{h}{4}.$
Il est facile de v\'erifier que $\boldsymbol \psi = \boldsymbol{\psi} ({\bf V}) $ est \`a divergence nulle dans $\Om$ et satisfait les conditions aux limites  $(\ref{BC})$.
\vskip -0,2 cm
\subsection{La Filtration}
\vskip -0,1 cm

On suppose que  ${\bf V} \in H^3({\cal O})^2$ et est \`a support compact, $\alpha>0$ est fix\'e, $\vit \in {\bf I \!  H}_1$. Soit
\BEQ \label{Filter}
\left \lbrace
\begin{array}{l}
-\alpha^2 \Delta \hat {\overline {\vit}} +
\hat {\overline {\vit}} + \g r =  \vit - \boldsymbol{\psi} ({\bf V}), \quad
\g \cdot \hat {\overline {\vit}} = 0, \\
\hat {\overline {w}} \vert _\Gamma = 0, \quad
\hat {\overline {\vit}}_h \vert_ {\Gamma_l \cup \Gamma_b }Ê= 0,
\quad \g \hat {\overline {\vit}}_h  \cdot {\bf n} \vert _{\Gamma_s} = {\bf 0}. \end{array}
\right.
\EEQ
En raisonnant par sym\'etrisation on montre dans \cite{Le08} que $(\ref{Filter})$ admet une unique solution $(\hat {\overline {\vit}}, r)$ avec $\hat {\overline {\vit}}  \in H^{3-\E} (\Om)^3$ pour tout $\E >0$. Soit $\overline \vit$ d\'efini par
$ \overline \vit = \hat {\overline {\vit}} + \boldsymbol{\psi} ({\bf V})$.
On note que
$\overline \vit$ a une trace sur $\G_s$ ainsi que $\g \overline \vit$ et donc  $\overline \vit$ satisfait $(\ref{BC})$. Soit $\vit = \vit (t, \x)$ \`a divergence nulle. L'\'equation de d\'econvolution est, o\`u $\tau$ est le param\`etre de d\'econvolution et $\vittest = \vittest(\tau, t, \x) = (\vittest_h, \theta)$, $t \in [0,T]$ est fix\'e,
\BEQ \label {EQSS} \left \lbrace  \begin{array} {l}
\displaystyle -\alpha^2 \Delta {\p \vittest \over \p \tau } + \vittest + \g \pi= \vit- \boldsymbol{\psi} ({\bf V}) = \hat \vit, \quad
\g \cdot \vittest = 0, \\ \displaystyle
\vittest_h \vert_{\Gamma_l \cup \Gamma_b}Ê= {\bf 0}, \quad
\g \vittest_h   \cdot {\bf n}\vert _{\Gamma_s}  = {\bf 0}, \quad \theta \vert _{\p \Om} = 0, \quad
\vittest (0, \x) = \overline \vit +  \boldsymbol{\psi} ({\bf V}) =
\hat {\overline \vit}.  \end{array}\right.  \EEQ
Soit $\tau >0$, $H_\tau (\vit)(t, \x) =
\vittest (\tau, t, \x) + \boldsymbol{\psi} ({\bf V})(\x)$. Avec la base
du Lemme \ref{BSP}, on \'ecrit pour
$\vit \in L^2([0,T], {\bf I \!  H}_1 ) \cap  L^\infty([0,T], {\bf I \!  H}_0 )  $,  $\vit  =  \sum_{k= 1}^\infty u_k (t) {\bf e}_k$ et
$\boldsymbol{\psi} ({\bf V}) =  \sum_{k= 1}^\infty \psi_k {\bf e}_k$. On a
\BEQ \label{FORM} H_\tau (\vit)(t, \x)  = \vit (t, \x) + \sum_{k= 1}^\infty (\overline u_k(t) - u_k(t) ) \exp (-{ \tau \over \alpha^2 \lambda_k})\,  {\bf e}_k (\x) , \, \,
\overline u_k(t) = {u_k (t) + \alpha^2 \psi_k \over 1 + \alpha^2 \lambda_k},\EEQ

\subsection{Le mod\`ele de d\'econvolution}

Le mod\`ele de SGE que nous consid\'erons est le suivant, avec $\vit$ la vitesse et $p$ la pression :
\begin{equation}
\label{pb1}\left\lbrace
\begin{array} {l}Ê
\partial_{t}\vit+ (H_{\tau} (\vit) \cdot \nabla) \, \vit-\nu\triangle\vit+\nabla
p={\bf f},\quad
\g \cdot \vit = 0, \\
w \vert _\Gamma = 0, \quad {\vit_h} \vert_ {\Gamma_l \cup \Gamma_b }Ê= 0,
\quad \g {\vit_h}  \cdot {\bf n} \vert _{\Gamma_s} = {\bf V}, \quad
\vit_{t=0} = \vit_0,  \end{array}
\right.
\end{equation}
Notons que quand $\tau=0$, il s'agit du Leray-$\alpha$ mod\`ele (\cite{CHOT05}). Nous prouvons dans \cite{Le08} :
\begin{theorem}Ê\label{Beau} Soit ${\bf V} \in H^3({\cal O})^2$ \`a support compact, $\vit_0  \in {\bf I \!  H}_1 $ et
${\bf f} \in
C^0([0,T],  {\bf I \!  H}_1)$.
Pour $\tau >0$ fix\'e, le syst\`eme $(\ref{pb1})$ admet une unique solution
$(\vit_\tau, p_\tau) \in L^2([0,T], H^2(\Om)^3)$
$\times$ $ L^\infty([0,T], H^1(\Om))$ avec $\vit_\tau = \vit_\tau (t, \x) \in C([0,T], L^2(\Om)^3)$,
$\p_t \vit_\tau \in L^2 ([0,T] \times \Om)^3$. \end{theorem}

\begin{theorem} \label{FINAL}  Il existe une suite $(\tau_n)_{n \in N}$ qui tend vers l'infini quand $n$ tend vers l'infini, telle que  $(\vit_{\tau_n})_{n \in  \N}$ converge faiblement dans
$ L^2([0,T], {\bf I \!  H}_1)$, fortement dans  $ L^2([0,T], {\bf I \!  H}_0)$, vers un champ
$\vit = (\vit_h, w) \in L^2([0,T], {\bf I \!  H}_1) \cap L^\infty([0,T], L^2(\Om)^3)$ avec
 $\p_t \vit \in L^{4 \over 3}Ê([0,T],  {\bf I \!  H}_1')$ et tel que pour chaque
$\vittest = (\vittest_h, \theta) \in L^4([0,T], {\bf I \!  H}_1)$ on a
\BEQ \label{varfaib} < \p_t \vit , \vittest > - \int_0^T \int_\Om \vit \otimes \vit :
\g \vittest + \nu\int_0^T \int_\Om \g \vit : \g \vittest -  \int_0^T \int_{\Gamma_s}Ê
{\bf V}Ê\cdot \vittest_h = \int_0^T \int_\Om {\bf f} \cdot \vittest.
\EEQ
De plus, $\vit (t, \cdot)$ tend faiblement vers $\vit_0$ dans $L^2(\Om)^3$ quand $t$ tend vers 0, et v\'erifie l'in\'egalit\'e d'\'energie.
\end{theorem}

\section{R\'esultats num\'eriques}

On r\'esoud $(\ref{pb1})$ par la m\'ethode des \'el\'ements finis en utilisant FreeFem++\cite{He06} dans le cas 2D, pour $\tau$ fix\'e.  On d\'etermine la pression et la vitesse par un sch\'ema semi-implicite initialis\'e avec des solutions d'un probl\`eme de Stokes. On a \'ecrit un algorithme de r\'esolution it\'eratif qui adapte la m\'ethode des caract\'eristiques au cas du transport d\'econvol\'e. On arr\^ete la simulation lorsque les algorithmes ont converg\'e. On pr\'esente un premier cas servant \`a valider notre mod\`ele de d\'econvolution et un deuxi\`eme cas qui montre son int\'er\^et pratique.

\subsection{Turbulence induite par le vent de surface}

On consid\`ere  $\Om = [0,1]\times[-0.5,0]$, triangul\'e  avec $h_{max}=0.014\hskip2pt m$ et $h_{min}=0.006\hskip2pt m$. On choisit ${\bf V}(x)=a(x)\sin(\pi x)$  o\`u $a(x)$ est une fonction de localisation \`a support compact. On compare les r\'esultats des simulations DNS et LES pour des param\`etres de d\'econvoution successivement \'egaux \`a $\tau=0$, $\tau=5$ ou $ \tau = 20$ gr\^ace \`a des profils verticaux ; ici $\Delta \tau=1$, $\alpha=0.1\, m$, $\Delta t = 0.2 \, s$ et on a fait $450$ it\'erations en temps, ce qui nous donne le r\'esultat pour $T=90 \, s$. On trouve une erreur $L_{2}$ moyenne de $0.5\hskip2pt\%$ entre les r\'esultats DNS et les r\'esultats du mod\`ele de d\'econvolution pour $\tau=20$ et de $1\%$ pour $\tau=5$, ce qui est satisfaisant et ce qui valide notre mod\'elisation. De plus, on a divis\'e l'erreur $L_{2}$ moyenne d'un facteur 4 entre les r\'esultats du Leray-$\alpha$ mod\`ele ($\tau = 0$) et ceux du mod\`ele de d\'econvolution pour $\tau=20$, ce qui montre l'am\'elioration apport\'ee.

\subsection{Cas avec bathym\'etrie}

On consid\`ere le domaine $\Om$ d\'ecrit ci-dessous avec une bosse, triangul\'e avec $h_{max}=0.010\hskip2pt m$ et $h_{min}=0.003\hskip2pt m$.

\newpage

On prend le m\^eme vent \`a la surface que pr\'ec\'edemment et on impose un flot de poiseuille sur les bords d'entr\'ee et de sortie $\Gamma_{l_{1}}$ et $\Gamma_{l_{2}}$ afin d'augmenter le niveau de turbulence. On conserve les param\`etres $\alpha=0.1 \, m$, $\Delta t = 0.02 \, s$, $\tau=5$ et on fait 500 it\'erations en temps, soit $T=10\, s$. On observe que la DNS converge difficilement avec des r\'esidus \'elev\'es, ce qui n'est pas le cas des mod\`eles LES (voir Figure 3.2).

\figgauche{8cm}{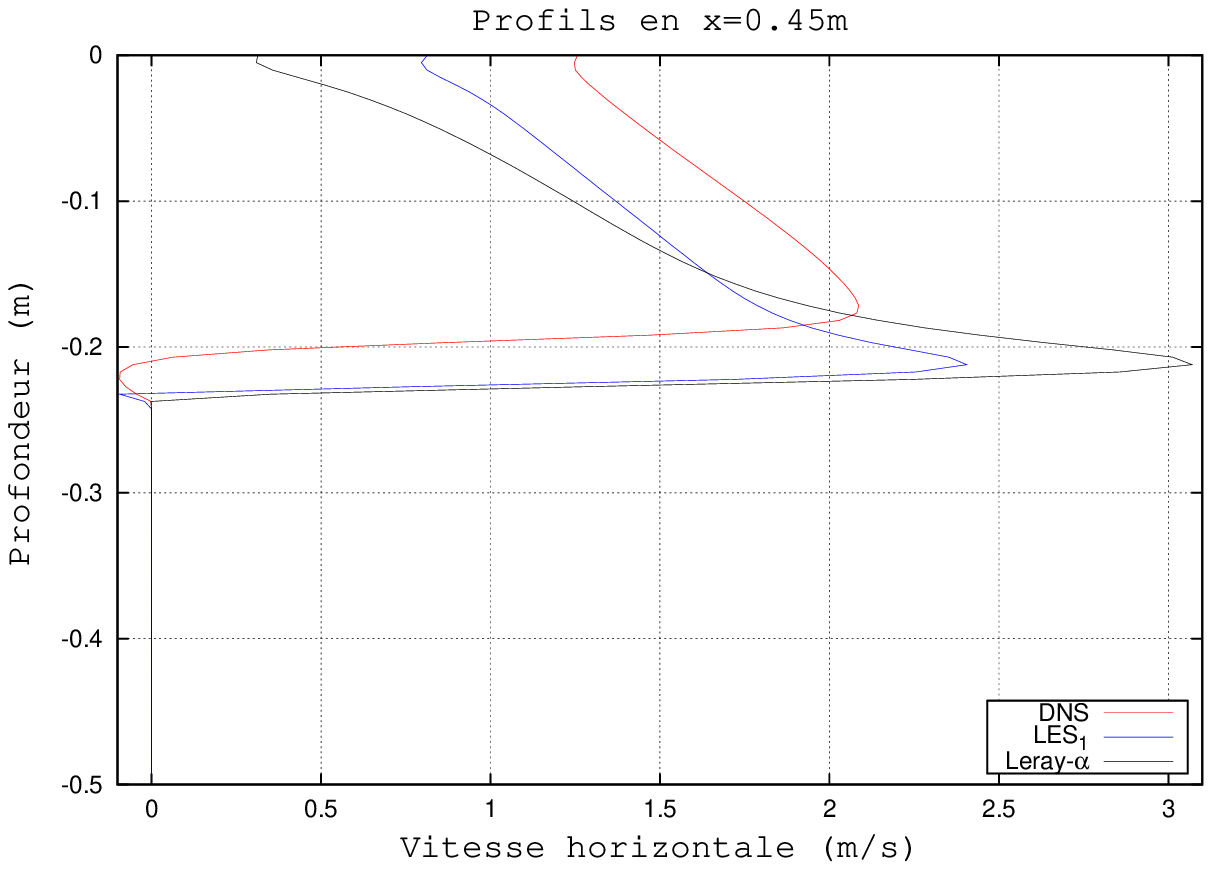}{decoll}{8cm}{On observe que notre mod\`ele am\'eliore la solution donn\'ee par le mod\`ele Leray-$\alpha$, ce qui est en accord avec les conclusions pr\'ec\'edentes. Ce profil montre qu'il faut augmenter le nombre de d\'econvolution qui est \'egal \`a 5 comme pr\'econis\'e par Adams et al \cite{St01}. Il est aussi n\'ecessaire d'abaisser la valeur du param\`etre de filtration $\alpha$, afin d'avoir un temps de calcul raisonnable par un abaissement du taux de d\'econvolution. Ce cas montre que notre mod\`ele peut corriger les effets de la filtration et montre aussi que le calage des param\`etres de d\'econvolution et de filtration n'est pas le m\^eme selon les cas \'etudi\'es. }

{\bf Aknowledgement}. The work of Roger Lewandowski is partially supported by the ANR project 08FA300-01. The work of Edriss S. Titi is partially supported by NSF grant No DMS-0708832 and the ISF Grant No 120/6.


\bibliographystyle{siam} \bibliography{Bib}

\begin{thebibliography}{1}

\bibitem{Le08}
{\sc A.-C. Bennis, R.~Lewandowski, and E.~S. Titi}, {\em A generalized
  {L}eray-deconvolution model of turbulence}, In Preparation,  (2008).

\bibitem{CHOT05}
{\sc A.~{C}heskidov, D.~D. Holm, E.~Olson, and E.~S. Titi}, {\em On a
  {L}eray-$\alpha$ model of turbulence}, Royal Society London, Proceedings,
  Series A, Mathematical, Physical and Engineering Sciences, 461 (2005),
  pp.~629--649.

\bibitem{GH03}
{\sc B.~J. Geurts and D.~D. Holm}, {\em Regularization modeling for large eddy
  simulation}, Physics of fluids, 15 (2003), pp.~L13--L16.

\bibitem{He06}
{\sc F.~Hecht, O.~Pironneau, A.~L. Hyaric, and K.~Ohtsua}, {\em Freefem++
  manual v2.21},  (2006).

\bibitem{La08}
{\sc W.~Layton and R.~Lewandowski}, {\em A high accuracy {L}eray-deconvolution
  model of turbulence and its limiting behavior}, Analysis and Applications, 6
  (2008), pp.~1--27.

\bibitem{St01}
{\sc S.~Stolz, N.~A. Adams, and L.~Kleiser}, {\em An approximate deconvolution
  model for large-eddy simulation with application to incompressible
  wall-bounded flows}, Physics of fluids, 13 (2001), pp.~997--1015.

\bibitem{MZ952}
{\sc M.~Ziane}, {\em Regularity results for a {S}tokes type system},
  {A}pplicable {A}analysis, 58 (1995), pp.~263--293.

\end{thebibliography}

\end{document}